\documentclass[pra,english,a4paper,preprintnumbers,apj,aps,superscriptaddress,twocolumn,floatfix,showkeys,showpacs]{emulateapj}

\usepackage{latexsym}
\usepackage{epsfig}
\usepackage{amssymb}
\usepackage{graphicx}

\newcommand{\ba}{\begin{eqnarray}}
\newcommand{\ea}{\end{eqnarray}}
\newcommand{\be}{\begin{equation}}
\newcommand{\ee}{\end{equation}}
\newcommand{\bk}{{\bar k}}

\newcommand{\br}{{\bf r}}

\newcommand{\R}{\mathcal{R}}
\newcommand{\A}{\mathcal{A}}
\newcommand{\lp}{\left(}
\newcommand{\rp}{\right)}
\newcommand{\lb}{\left[}
\newcommand{\rb}{\right]}
\newcommand{\rr}{\hat{r}}

\def\beq{\begin{equation}}
\def\eeq{\end{equation}}
\def\baq{\begin{eqnarray}}
\def\eaq{\end{eqnarray}}

\def\k{{\bar k}}
\def\q{{\bar q}}
\def\dq{{\rm d}{\bar q}}
\def\x{{\bar x}}

\def\lnkkp{{\rm ln}\,\frac{k}{k_{\rm p}}}

\def\fnlth{f_{{\rm NL},\theta}}
\def\nfnl{n_{f_{{\rm NL},0}}}

\begin{document}
\title{CMB statistics in noncommutative inflation}

\author{Tomi S. Koivisto$^1$ and David F. Mota$^2$}
\affil{$^1$ITF and the Spinoza Institute, Postbus 80.195, 3508 TD Utrecht, The Netherlands.}
\affil{$^2$Institute of Theoretical Astrophysics, University of
Oslo, 0315 Oslo, Norway}

\pacs{98.80.-k,98.80.Cq,11.10.Nx.}
\preprint{SPIN-10/34}
\preprint{ITP-UU-10-41}
\keywords{Cosmological inflation, Statistical anisotropy of the cosmic microwave background, Scale-dependent non-Gaussianity, Noncommutative spacetime}

\begin{abstract}

Noncommutative geometry can provide effective description of physics at very short distances taking into account generic effects of quantum gravity. Inflation amplifies tiny quantum fluctuations in the early universe to macroscopic scales and may thus imprint high energy physics signatures in the cosmological perturbations that could be detected in the CMB. It is shown here that this can give rise to parity-violating modulations of the primordial spectrum and odd non-Gaussian signatures. The breaking of rotational invariance of the CMB provides constraints on the scale of noncommutativity that are competitive with the existing noncosmological bounds, and could explain the curious hemispherical asymmetry that has been claimed to be observed in the sky. This introduces also non-Gaussianity with peculiar shape- and scale-dependence, which in principle allows an independent cross-check of the presence of noncommutativity at inflation.


\end{abstract}

\section{Introduction}

The statistics of the temperature anisotropies in the cosmic microwave background (CMB) are measured by the Planck satellite to an unprecedented accuracy. This allows to
efficiently probe, in addition to higher-order correlations, i.e. possible non-Gaussianity, the detailed structure of the two-point correlations, i.e. possible statistical anisotropy.

There is a number of anomalies already in the present data, which have raised a lot curiosity both from the theoretical side as well as from the data analysis side \citep{Hansen:2008ym,Hoftuft:2009rq,Koivisto:2005,Koivisto:2007,Koivisto:2008,Rakic:2006tp,Koivisto:2009,Koivisto:2010,Koivisto:2007b,Rakic:2007ve,Koivisto:2008b,Koivisto:2009b,Koivisto:2010b,
Copi:2010na,Bennett:2010jb}. In particular, the hemispherical asymmetry, first reported by \cite{Eriksen:2003db}, seems a quite unexpected feature within the standard model of cosmology and hasn't yet been a satisfactorily traced to a possible systematic error.
The question whether the universe is odd was asked in \citep{Land:2005jq}, and there are recent investigations \citep{Kim:2010gf,Kim:2010st} finding hints of evidence for a positive answer.

This prompts to look for possible cosmological origins of odd-parity statistical anisotropies. In the present study, we investigate the effects of noncommutative geometry to the primordial spectrum of perturbations, usually assumed to be generated by quantum effects during inflation or shortly afterwards. The observational implications are derived, in terms of the harmonic coefficients of the CMB spectrum, and the non-Gaussianity parameter $f_{NL}$. We find that in general the noncommutativity of spacetime geometry induces parity violating modulations of the spectra of fluctuations, thus generating distinct signatures in the statistics of CMB. In particular, this suggests that the hemispherical asymmetry (and various other anomalies) could originate from the fundamental properties of spacetime that are relevant at the vast energy scales at play in the inflationary epoch.

\cite{Chu:2000ww} remarked that spacetime noncommutativity can be constrained by the statistics of inflationary fluctuations.
The power spectrum has been computed \citep{Tsujikawa:2003gh,Koh:2007rx} and various other aspects of noncommutative inflation have been discussed in the literature \citep{Alexander:2001dr,Calcagni:2004ug,Palma:2009hs}. The CMB constraints beyond the power spectrum have been explored also \citep{Lizzi:2002ib,Akofor:2007fv,Akofor:2008gv,Karwan:2009ic}. Here will adopt the formalism of \cite{Akofor:2007fv}. What is new in particular, is that we point out the presence of odd signatures and compute the structure of the two-point and three-point correlators in more detail and generality than previously.

We also clarify an ambiguity of the results, which forces us to introduce an additional parameter. To assess the robustness of the results, we consider in addition the alternative approach of \cite{Kobakhidze:2008cq}. It becomes clear that the details of the predictions can depend upon the particular model, but there are generic features which appear already in the simplest cases (in particular, in the case of canonical noncommutativity with constant $\theta$ in the comoving frame). These nontrivial statistical features may thus be present, at an observable level, even in the simplest inflationary $\Lambda$CDM models, if one takes into account the effect of spacetime uncertainty principle on the inflationary fluctuations.

However, other means of generating parity violations can be introduced too. A simple way is to assume an inhomogeneity present at the early universe. \citep{Erickcek:2009at} considered that
a large scale perturbation of the curvaton field might result in a power asymmetry. \cite{Tangen:2009mw} calculated the CMB pattern from a single superhorizon perturbation, which indeed shows couplings between adjacent multipoles. This is different from the approach of considering dipole in the primordial spectrum, which introduces adjacent-mode correlations for the anisotropies of the random fluctuations at all scales, as will become clear below. One may also contemplate on possible parity-violating couplings of the inflaton field. \cite{Alexander:2006mt} has considered the possible role of Chern-Simons terms \citep{Chern:1974ft}. Finally, spontaneously broken isotropy, occurring due to imperfect dark energy, has been shown to produce odd modulations \citep{Gordon:2005ai,magnus}. There are qualitative differences to the present case, which will be clarified in section \ref{comparison}.

In the following section \ref{noncommutative_perturbation} we review the basic results of inflationary perturbations and discuss how these can be applied when the spacetime is noncommutative. We then implement this in section \ref{geometry} in the case of canonically deformed spacetime commutation relations and in section \ref{nc_fields} in a framework based on deformed Heisenberg algebra of quantum fields. We are then ready to discuss the observable patterns in the CMB sky. The properties of the two-point functions and of the non-Gaussianities are clarified in section \ref{patterns}, and section \ref{conclusions} is a brief conclusion. The CMB two-point correlation in terms of the multipole expansion of the primordial spectrum is given in the section \ref{app_mul}.


\label{noncommutative_perturbation}

\section{Curvature perturbation in non-commutative inflation}

In the vast majority of models, primordial perturbations originate from quantum fluctuations of light scalar fields
produced by the inflationary expansion. Their properties depend on the physics operating at the very high energy scales
present during inflation. It is conceivable that physics at such high energies becomes inherently non-local; such models
can be effectively described by noncommutative theories. Here we are interested in studying primordial perturbations
generated in noncommutative theories of inflation. We treat gravity as a classical background which is not affected by
the noncommutative effects. In this approach the noncommutativity affects only the properties of quantum fluctuations
generated during inflation.

To keep the discussion transparent, we restrict our analysis
on general single field models where primordial perturbations effectively arise from fluctuations of a single scalar degree
of freedom $\phi$ while additional scalars may affect the background evolution. This class of models obviously contains
the standard single field inflation in which case $\phi$ is the inflaton. In general, however, $\phi$ can be a scalar field
different from the inflaton-like fields which dominate the energy density. Well known examples are the curvaton
model and modulated reheating scenario where primordial perturbations can arise solely from fluctuations of a light
field $\phi$ which remains subdominant during inflation but affects the expansion history at a later stage \citep{Lyth:2002my,Dvali:2003em}.

The primordial perturbations are conveniently characterized by the curvature perturbation $\zeta$ which measures fluctuations in the
spatial curvature on uniform energy density hypersurfaces. Since we take gravity as a classical background which is not
affected by the noncommutative effects, the curvature perturbation can be computed using the $\delta N$ formalism in close
analogue to the standard commuting case. On superhorizon scales we can write
  \beq
  \label{deltaN}
  \zeta_{\theta}(t,\x)=N'(t,t_i)\delta\phi_{\theta}(t_i,\x)+\frac{1}{2}N''(t,t_i)\delta\phi_{\theta}(t_i,\x)^2+\ldots \ ,
  \eeq
where the subscript $\theta$ is introduced to denote noncommutative variables. The scalar field perturbations
$\delta\phi_{\theta}$ generated during inflation are evaluated on a uniform curvature hypersurface $t_i$ soon after the
horizon crossing of all the modes of interest. Their properties differ from the corresponding commuting quantities
$\delta\phi_{0}$ as we will discuss below. The function $N(t,t_i)$ measures the number
of e-foldings of a classical Friedmann-Robertson-Walker (FRW) universe from the uniform curvature hypersurface at $t_i$ to a uniform density
hypersurface at some final time $t$ when the universe is evolving adiabatically. The primes denote derivatives with
respect to the classical background field $\phi$. The derivatives of $N(t,t_i)$
describe entirely classical properties of the theory and their values
coincide with the corresponding commutative theory. The curvature perturbations produced in a noncommutative and
commutative theory with the same classical solutions therefore differ only by the different properties of
$\delta\phi_{\theta}$ and $\delta\phi_{0}$. We turn to discuss the relation between
$\delta\phi_{\theta}$ and $\delta\phi_{0}$ in more detail after briefly reviewing some standard results for $\delta\phi_{0}$.

\subsection{The mode functions in the commutative case}
\label{inflation}


We consider the FRW metric in terms of conformal time $\tau$ and including scalar perturbations in the Newtonian gauge in the absence of shear \citep{Mukhanov:2007zzc}:
\be
ds^2 = -a^2(\tau)\lb d\tau^2\lp 1+2\Phi\rp - dx^2\lp1+2\Phi\rp\rb\,.
\ee
A scalar field $\psi$ can be expanded in terms of the annihilation and creation operators as
\be
\psi(x) = \int \lp u_k(\tau)e^{i{\bf k}\cdot{\bf x}} a_k + u^*_k(\tau)e^{-i{\bf k}\cdot{\bf x}} a^\dagger_k \rp \frac{d^3{\bf k}}{a(\tau)(2\pi)^3}\,.
\ee
The operators satisfy the canonical commutation relations. The canonical momentum can then be identified as
\be
\pi(x) = \frac{d}{d\tau}\lp a(\tau)\psi(x)\rp\ \equiv (a\psi)',.
\ee
mode functions $u(\tau)$ obey the time evolution equation
\be
u_k'' + \lp k^2-\frac{a''}{a}\rp u_k = 0\,,
\ee
with the well known Hankel function solutions that, when matched with the initial Bunch-Davies vacuum at early times, reduce at late times outside the horizon to
\ba \label{modes}
u_k(\tau) & = & \frac{e^{-ik(\tau-\tau_i)}}{\sqrt{2k}}\lp\frac{i}{k\tau}-1\rp\,, \\
u_k'(\tau) & = & \frac{ike^{-ik(\tau-\tau_i)}}{\sqrt{2k}}\lp 1-\frac{i}{k\tau}-\frac{1}{k^2\tau^2}\rp\,.
\label{modes2}
\ea
The spectrum of scalar metric perturbation $\Phi$ in the conformal Newtonian gauge is then related to the spectrum of the field fluctuation as
\be \label{p_phi}
P_\Phi(k) = \frac{16\pi G}{9\epsilon}P_\psi(k)\,,
\ee
and evaluated at the horizon crossing $a(\eta)H=k$, where the Hubble rate is approximately constant when the slow-roll parameter
\be
\epsilon \equiv -\frac{\dot{H}}{H^2}\,,
\ee
is small.
The spectrum of the metric perturbation $\Phi_0$ in standard single field inflation can be known to be given by
\be \label{p_0}
P_{\Phi_0}(t,k) = \frac{8\pi G}{9\epsilon k^3a^2(t)\tau^2(t)}\,.
\ee

\section{Canonical noncommutativity}
\label{geometry}

A canonical way of deforming the spacetime is to introduce the commutation relations for the coordinate operators
\be \label{com}
\lb \hat{x}_\mu, \hat{x}_\nu \rb = i\theta_{\mu\nu}\,,
\ee
where in the simplest case $\theta_{\mu\nu}$ is an antisymmetric constant matrix of dimension length squared. It is well known that this is the exact low-energy limit of open string theory with a constant antisymmetric background field \citep{Seiberg:1999vs}. In general, a commutation relation of the form (\ref{com}) induces the uncertainty relation for coordinates
\be
\Delta x^\mu \Delta x^\nu \ge \frac{1}{2}|\theta^{\mu\nu}|\,,
\ee
so that a spacetime point is heuristically replaced by a Planck cell. The ordinary coordinates may then be thought to be obtained by coarse-graining over scales smaller than the fundamental scale of order $\sqrt{|\theta|}$\,. Thus, noncommutative spacetime provides a framework that is compatible with generic features of quantum gravity like the uncertainty principle and nonlocality.

The commutation relations (\ref{com}) generally assumes more complicated form when expressed in alternative coordinate systems. It is thereby essential to specify in which frame this relation is taken to hold as written down above. In cosmology, a natural frame to consider is the comoving one. We call the physical scale $\theta^{ph}_{\mu\nu}$, whereas $\theta_{\mu\nu}$ is the matrix corresponding to the coordinates of an observer, to whom this matrix then is a constant throughout the evolution of the universe. We perform the computations in the comoving frame, but in the end translate the result into the physical scale employing the relations
\be \label{cm_ph}
\theta^{ph}_{0i}=a(t)\theta_{0i}(t)\,, \quad \theta^{ph}_{ij}=a^2(t)\theta_{0i}(t)\,,
\ee
where the $\theta_{\mu\nu}$ here and in the following is evaluated in the comoving coordinates. 

Consistent statistics in noncommutative spacetime \cite{Akofor:2008ae} require deformation of the quantum fields by the exponential operator defined by the following relation:
\be \label{fields}
\varphi_\theta(x) = \varphi_0(x) \exp{\lp-\frac{i}{2}\overleftarrow{\partial}_\mu \theta^{\mu\nu}\overrightarrow\partial_\nu\rp}\,,
\ee
where the lower index $\theta$ refers to the deformation, so $\varphi_0$ is the corresponding field in the commutative case.
In the following we will be interested in the two-point correlation of the inflaton field in this setting. This implies that the vacuum expectation value of the two-point function can now be written as
\be \label{two-point}
\langle 0| \varphi^\dagger_\theta(x) \varphi_\theta(x')|0 \rangle =
e^{-\frac{i}{2}{\partial}_{\mu} \theta^{\mu\nu}\partial_{\nu'}}
\langle 0| \varphi_0(x) \varphi_0(x')|0 \rangle\,.
\ee
Writing this in terms of the Fourier image $\phi_\theta({\bf k},t)$ defined by
\be
\varphi_\theta(x) = \int \frac{d^3k}{(2\pi)^3}\phi({\bf k},t)e^{i{\bf k}\cdot{\bf x}}\,,
\ee
we obtain
\ba \label{corr}
\langle 0| \varphi_\theta(x)^\dagger \varphi_\theta(x')|0 \rangle = \qquad\qquad\qquad\qquad\qquad\qquad\qquad\qquad\\ \nonumber  \int  \frac{d^3k}{(2\pi)^3}
e^{-\frac{i}{2}\lp k_i\theta^{i0}\partial_{t'} - \partial_t\theta^{0i}k_i\rp}
\langle 0|\phi^\dagger_0({\bf k},t)\phi_0({\bf k},t')|0 \rangle e^{i{\bf k}\cdot{\bf x}}
\ea
where we have used the fact that in the usual case $\theta=0$ (only) the different wavemodes are uncorrelated,
\be
\langle 0| \phi^\dagger_0({\bf k},t) \phi_0({\bf k'},t)|0 \rangle = (2\pi)^3 P_{\phi_0}(k,t)\delta^3({\bf k}-{\bf k'})\,,
\ee
where $P_\varphi(k,t)$ is the power spectrum. Let us call the time-space components of the noncommutativity the three-vector $\vec{\theta}$ as
\be
\theta^{0i} \equiv \vec{\theta}^i\,.
\ee
By comparing the form (\ref{corr}) to the usual case we readily infer that
\ba \label{p_theta}
\langle 0|\phi^\dagger_\theta({\bf k},t)\phi_\theta({\bf k},t)|0 \rangle  & = & \\ \lim_{t\rightarrow t'}
e^{-\frac{1}{2}\vec{\theta}\cdot {\bf k}(\partial_t + \partial_{t'})} &\times&\langle 0|\phi^\dagger_0({\bf k},t)\phi_0({\bf k},t')|0\rangle \nonumber \\
& = &  (2\pi)^3 P_{\phi_0}\lp k,t-\frac{1}{2}\vec{\theta}\cdot {\bf k}\rp \,.\nonumber
\ea
Consider then the spectrum of metric perturbation (\ref{p_0}) in near de Sitter space where $H$ is approximately constant is given by
\be
\tau \simeq \frac{-1}{Ha(t)}e^{-H t}\,.
\ee
Using this we can immediately combine equations (\ref{p_theta}) and (\ref{p_0}) to obtain the spectrum in noncommutative geometry. Evaluated at the horizon crossing, we have
\be \label{geometry_p}
P_{\Phi_\theta}({\bf k})  =  P_{\Phi_0}(k)e^{H  \vec{\theta}\cdot {\bf k}}\,.
\ee
Thus the spectrum will be direction-dependent. Furthermore, it is not parity invariant. We see that the leading correction is a dipole with an amplitude $\A_{1m} \sim |\vec{\theta}|$ and blue-tilted spectral index $n_{1,m} \simeq 2$. The next correction is the even-parity quadropole term, with an amplitude
$\A_{2,m} \sim -|\vec{\theta}|^2$ and spectral index $n_{1,m} \simeq 3$, and so on.

However, we have to choose the consistent parts of the correlator in order to obtain a physical result.
As clear from the subsection \ref{app_mul}, the odd multipole modulations should have imaginary coefficients, otherwise the result is not sensible as the real-space correlators and the CMB sky would not turn out real. This would be cured if we promote $\vec{\theta}$ into an imaginary parameter but this seems inconsistent with \ref{com}. Therefore we adopt the following  prescription
\be \label{pres}
\langle\dots \rangle \rightarrow \alpha\langle\dots\rangle_M + i(1-\alpha)\langle\dots\rangle_A\,,
\ee
where $\langle\dots\rangle$ denotes schematically some a correlator, $\langle\dots\rangle_M$ is its self-adjoint,
and $\langle\dots\rangle_A$ its anti-self-adjoint part. Then $\alpha \in [0, 1]$ is a parameter which corresponds to some kind
of phase. \cite{Akofor:2007fv} considered only the self-adjoint part of the correlators, which corresponds to $\alpha=1$ in our parametrization. However, we do not know any physical reason why the remaining part should not contribute to observed correlations.
As we are unable to determine the value of $\alpha$ from first principles, it is left as a parameter to be determined by observations. We then introduce the notation
\be \label{expa}
\exp_\alpha(x) \equiv \alpha \cosh{(x)} + i(1-\alpha)\sinh{(x)}\,.
\ee
In this prescription, the result (\ref{geometry_p}) becomes
\be \label{geometry_p2}
P_{\Phi_\theta}({\bf k})  =  P_{\Phi_0}(k)\exp_\alpha(H  \vec{\theta}\cdot {\bf k})\,,
\ee
which is the main result of this section.

The conceptual problem noted above appears in alternative frameworks too and is thus not merely a possible inconsistency in of the particular formalism employed here. In particular, one could also start by expanding the action in the noncommutative parameter and derive the correlations of the perturbations from the ensuing equations of motion. One obtains also then imaginary results, see e.g. the four-point function calculated by \cite{Chu:2000ww}. It can be pointed out also that the problematics of observing correlations of noncommuting observables are independent of the nature of noncommutativity.
Noncommutativity between space and time has not yet been put into theoretically rigorous footing but generically
seems to imply unitarity violations \citep{Gomis:2000zz,Chaichian:2000ia}. Though in open string theory with a constant electric background, which is supposed to exhibit noncommutativity between space and time, these problems are absent \citep{Seiberg:2000gc}, the noncommuting field theory with constant $\theta_{\mu\nu}$ can be recovered from string theory only in the case of magnetic background field which then corresponds to vanishing $\vec{\theta}$. The need for the prescription (\ref{pres}) would nevertheless reappear for higher order correlations even when $\vec{\theta}=0$ as will be seen in section \ref{nong}.

\section{Deformation of the Heisenberg algebra}
\label{nc_fields}

Violation of microcausality in the spirit of (stringy) uncertainty principle can also be described by imposing noncommutativity of quantum fields \citep{Dubovsky:2007ac} (instead of the coordinate operators, as in the previous subsection). Consider the following equal-time commutation relations in expanding spacetime
\ba
\lb \phi({\bf x},\tau),\phi({\bf y},\tau) \rb & = & \frac{i\mu^2(\tau)}{a^2(\tau)}f({\bf x}-{\bf y})\,, \label{ncf1}\\
\lb \pi({\bf x},\tau),\pi({\bf y},\tau)   \rb  & = & 0\,, \\
\lb \phi({\bf x},\tau),\pi({\bf y},\tau) \rb & = & \frac{i}{a^2(\tau)}\delta({\bf x}-{\bf y}) \label{ncf3}\,.
\ea
For notational convenience we parametrize $\mu(\tau)=\mu_0a(\tau)^\frac{n}{2}$, where the constant where $\mu_0$ is the characteristic scale of microcausality violation with the dimension dim[$\phi$]-dim[$f$]/2. The scale-factor dependence is added because we want to consider also the case where the form of the commutator is constant in comoving coordinates, $n=2$. The time-dependence of the effective parameter $\mu$ does not affect the computation and we return to the different choices in the analysis of the results. The derivation here follows closely \cite{Kobakhidze:2008cq}, where $n=0$.
The difference to the usual case is now only the odd function $f({\bf r})$ appearing in the first commutator.
It is useful to note that by defining the field $\psi(x)$ as\footnote{The factor $1/a(\tau)$ was missing in Eq.(6) of \cite{Kobakhidze:2008cq}.}
\be \label{shift}
\phi(x) = \psi(x) - \frac{\mu^2(\tau)}{2a(\tau)}\int f({\bf x}-{\bf z})\pi({\bf z},\tau)d^3{\bf z}\,,
\ee
we recover canonical commutation relations for the pair $(\psi,\pi)$. This observation allows us, analogously to the previous subsection, to relate correlations in the noncommutative case to the correlators in the standard case. In particular, one may check that if the field $\psi$ satisfies the equations for standard inflaton we described in section \ref{inflation}, one can translate those conventional results into the noncommutative set-up (\ref{ncf1}-\ref{ncf3}) by employing the shift (\ref{shift}). The noncommutative inflaton field is then expanded is terms of the mode functions $u_k$ introduced in section \ref{inflation} as
\ba
\phi(x) & = & \int\frac{d^3{\bf k}}{(2\pi)^3}
\Bigg[ \lp u_k(\tau)-\frac{\mu^2(\tau)}{2}F({\bf k})u'_k(\tau)\rp e^{i{\bf k}\cdot{\bf x}} a_k \nonumber \\ & + &
\lp u^*_k(\tau)+\frac{\mu^2(\tau)}{2}F({\bf k}){u^*_k}'(\tau)\rp e^{-i{\bf k}\cdot{\bf x}} a^\dagger_k\Bigg]\frac{1}{a^2(\tau)}\,, \label{shifted}
\ea
where $F({\bf k})$ is the Fourier image of the function $f({\bf x})$,
\be
F({\bf k}) = \int f({\bf y}) e^{i{\bf k}\cdot{\bf y}}d^3{\bf y}\,.
\ee
Since $f(\bf{r})$ is an odd function, we have also $F(-{\bf k})=F({\bf k})$.
The two-point correlation function follows then straightforwardly:
\ba
\langle 0|\phi({\bf x},\tau)\phi({\bf y},\tau)|0\rangle  & = & \int \frac{d^3{\bf k}U_{\bf k}(\tau)U^*_{\bf -k}(\tau)e^{i{\bf k}\cdot ({\bf x}-{\bf y})}}{a^2(\tau)(2\pi)^3}
\nonumber \\
& = & \int P^\mu_\phi({\bf k}) e^{i{\bf k}\cdot ({\bf x}-{\bf y})} \frac{d^3{\bf k}}{(2\pi)^3}\,.\label{mu_s1}
\ea
where $U_\bk(\tau)$ is defined as
\be \label{u_nc}
U_{\bf k}(\tau) = u_k'(\tau)-\frac{\mu^2(\tau)}{2}F(\bk)u'_k(\tau)\,.
\ee
In the second line we have identified the power spectrum of the non-selfcommuting field $\phi$ and denoted it by $P^\mu_\phi({\bf k})$.
  It is easy to see that lest physical observables become imaginary, the Fourier image $F({\bf k})$ must be real, and thus we cannot express the first line of (\ref{mu_s1}) as a square\footnote{Our result (\ref{shifted}) differs from the Eq.(13) in \cite{Kobakhidze:2008cq} by the sign of the argument of the second $F({\bf k})$, but our (\ref{mu_s1}) would agree with the Eq.(14) in \cite{Kobakhidze:2008cq} if $F({\bf k})$ was imaginary. This indeed seems to have been the assumption in \cite{Kobakhidze:2008cq}, which then however leads to imaginary temperature correlations, see Eq.(23) there.}. Now using the relation of the inflaton and metric perturbation spectra (\ref{p_phi}) and the late-time limit of the mode function solutions (\ref{modes},\ref{modes2}) we obtain
\ba \label{p_phinc}
P^\mu_\Phi({\bf k}) & = & \frac{8\pi G}{9\epsilon k^3a^2\tau^2}\Big[ 1 + k^2\tau^2\lp 1 + i\mu^2 k F({\bf k}) \rp \nonumber \\
& - & \frac{\mu^4}{2\tau^2}F^2({\bf k})\lp 1 - k^2\tau^2 + k^4\tau^4 \rp \Big]\,.
\ea
The leading order contribution in the parameter is thus odd in parity.

Let us then look at some specific forms of the noncommutativity. A simple assumption for the form of the function $f({\bf r})$ is a
delta-function and that the commutator is constant in comoving coordinates. An odd combination is
\be \label{delta}
\mu^2(\tau)f({\bf r})  = 4\pi^3i \mu_0^2a^2\lb\delta({\bf r}-{\bf v})-\delta(-{\bf r}-{\bf v})\rb\,.
\ee
In this prescription, the commutator (\ref{ncf1}) gets contribution from spacelike separations equal to ${\bf v}$. Then (\ref{p_phinc}) becomes, neglecting the decaying modes
\ba \label{fields_p}
P^\mu_\Phi({\bf k}) & = & P_\Phi(k)\Big[ 1 + k^3(\mu_0 a\tau)^2 \sin({\bf k}\cdot{\bf v}) \nonumber \\
& + & \lp\frac{\mu_0a}{\sqrt{2}\tau}\rp^4\sin^2({\bf k}\cdot{\bf v}) \Big]\,.
\ea
Remarkably, at the leading order the predicted modulation has the same form as in the previous case (\ref{geometry_p}),
if we identify the vectors ${\bf v}=H\vec{\theta}$. One of course obtains hyperbolic sinus instead of the ordinary one by promoting ${\bf v}$ to an imaginary vector parameter (and dropping the $i$ in (\ref{delta})).
The second line in (\ref{fields_p}) contains modes growing outside the horizon. They can be eliminated by choosing $n \le -1$ in $\mu(\tau)=\mu_0a^n(\tau)$. If $n<-1$ both the odd and the even contributions are decaying (regardless of the form of $f(\br)$). The spectral index of the leading modification in (\ref{fields_p}) is $n_S+3$, but this depends sensitively on the precise form of $f(\br)$. As an example, the form
\ba
f(\br)& = & i\frac{\pi^3}{2}\Big[ (r_z-v_z)\sigma(v_x - r_x) \sigma(v_y - r_y)\sigma(v_z-r_z) \nonumber \\
& + & (r_z+v_z)\sigma(v_x + r_x) \sigma(v_y + r_y)\sigma(v_z + r_z)\Big]\,,
\ea
where $\sigma(x)$ is the sign of $x$, results in $F({\bk})=\sin({\bf k}\cdot {\bf v})/(k_xk_yk_z^2)$, which results in a strongly blue-tilted spectral index. Thus we may obtain similar correlations as in section \ref{geometry} by choosing a suitable function $f({\bf r})$.

For the three-point function one gets
\ba \label{tpc_h}
\langle \phi(\bk_1)\phi(\bk_2)\phi(\bk_3) \rangle & = &  \frac{U_{\bf k_1}(\tau)U_{\bf k_3}^*(\tau)}{\lp 2\pi a(\tau) \rp^6}  \\ \nonumber  \Big[ \delta^3({\bf k_1+k_2-k_3})U_{\bf k_2}(\tau) & + & \delta^3({\bf k_1-k_2-k_3})U_{\bf k_2}^*(\tau)\Big]\,.
\ea
in terms of the functions $U_{\bk}(\tau)$ defined in (\ref{u_nc}).

\section{Patterns from noncommutative inflation}
\label{patterns}

In this section we discuss some observational implications of the results at more length and derive an explicit expression for the non-Gaussianity.

\subsection{Multipole expansion of the primordial power spectrum}
\label{app_mul}

The temperature anisotropy field is conventionally expanded in terms of the spherical harmonics and on the other
hand considered in the Fourier space
\be \label{exps}
\Theta({\bf x}, \hat{e},\eta) = \sum_{\ell=0}^{\infty}\sum_{m=-\ell}^{\ell}a_{\ell m} Y_{\ell m} =
\int \frac{d^3k}{(2\pi)^3} e^{i{\bf k \cdot x}} \R_{\bf k} \Theta({\bf k},\hat{e}, \eta)\,,
\ee
where we have normalized the transfer function $\Theta({\bf k},{\bf e},\eta)$ with respect to
the initial amplitude of the primordial curvature perturbation $\R_{\bf k})$.
It follows that the coefficients $a_{\ell m}$ are given by
\be \label{alm}
a_{\ell m} = i^\ell \int \frac{d^3k}{2\pi^2}\R_{\bf k} Y^*_{\ell m}(\hat{k})\Theta_l(k).
\ee
where we have introduced the transfer function which depends only on the magnitude of the wavevector,
\be
\Theta_l(k) = \int j_\ell (kr(\eta))\Theta(k,\eta)d\eta\,,
\ee
since we assume the evolution to be isotropic. The possible anisotropy appears in the primordial spectrum, which can be expanded also in
spherical harmonics as (we refer to this expansion of the primordial spectra always with capital letters indices to avoid confusion with
the expansion of the temperature anisotropies)
\ba \label{primordial}
\langle \R_{\bf k} \R^*_{\bf k'} \rangle &=& \delta^3({\bf k}-{\bf
  k'})\frac{2\pi^2}{k^3}\mathcal{P}({\bf k}) = \\ \nonumber
\delta^3({\bf k}&-&{\bf k'})\frac{2\pi^2 \sqrt{4\pi}}{k^3}\sum_{L=1}^{\infty}\sum_{M=-L}^{M}\A_{L M}
\left(\frac{k}{k_0}\right)^{n_{L M}-1}Y_{L M}(\hat{k})\,,
\ea
In the first equality we have used the WMAP conventions, and in the second one employed the parametrization of \cite{ArmendarizPicon:2008yr}.
They use the pivot scale $k_0=2\cdot10^{-3}/$Mpc. For the time being, we allow independent spectral indices $n_{LM}$ for each
multipole $L$,$M$. Using the formula (\ref{alm}) and the primordial
spectrum (\ref{primordial}), the
correlation matrix can be written as
\be
\langle a_{\ell m} a^*_{\ell' m' } \rangle = \frac{i^{\ell-\ell'}}{2\pi^2}\sum_{L=1}^{\infty}\sum_{M=-L}^{L}\A_{L M}
\xi^{L M}_{\ell m;\ell' m'} I^{L M}_{\ell \ell'}\,.
\ee
We have separated here the integrated contribution from perturbations of all different magnitudes,
\be
I^{L M}_{\ell \ell'} = \int_{0}^{\infty}\frac{dk}{k}\left(\frac{k}{k_0}\right)^{n_{LM}-1}\Theta_\ell(k)\Theta_{\ell'}(k)\,.
\ee
They are then weighted by the geometrical factors $\xi$ which happen to be proportional to the coefficients of the Gaunt series.
We may write them in terms of the Wigners 3-functions as
\begin{widetext}
\be
\xi^{L M}_{\ell m;\ell' m'} = (-1)^{m+1}\sqrt{(2\ell+1)(2\ell'+1)(2L +1)}
\left(
\begin{array}{ccc}
\ell & \ell' & L \\
0 & 0 & 0 \\
\end{array} \right)
\left(
\begin{array}{ccc}
\ell & \ell' & L \\
-m & m' & M \\
\end{array} \right)\,,
\ee
when $\ell$, $\ell'$ and $L$ satisfy the triangle condition. For the dipole these become
\be \label{zeta+}
\xi^{L=1,M=-1}_{\ell m;\ell' m'} = \sqrt{3}\delta_{m',m-1}\left[
\delta_{\ell',\ell-1} \sqrt{\frac{{(\ell+m-1)(\ell+m)}}{2(2\ell-1)(2\ell+1)}} -
\delta_{\ell',\ell+1} \sqrt{\frac{{(\ell-m+1)(\ell-m+2)}}{2(2\ell+1)(2\ell+3)}} \right],
\ee
\be \label{zeta-}
\xi^{L=1,M=1}_{\ell m;\ell' m'} = \sqrt{3}\delta_{m',m+1}\left[
\delta_{\ell',\ell-1} \sqrt{\frac{{(\ell-m-1)(\ell-m)}}{2(2\ell-1)(2\ell+1)}} -
\delta_{\ell',\ell+1} \sqrt{\frac{{(\ell+m+1)(\ell+m+2)}}{2(2\ell+1)(2\ell+3)}} \right],
\ee
\be \label{zeta0}
\xi^{L=1,M=0}_{\ell m;\ell' m'} = \sqrt{3}\delta_{m',m}\left[
\delta_{\ell',\ell-1} \sqrt{\frac{{(\ell-m)(\ell+m)}}{(2\ell-1)(2\ell+1)}} +
\delta_{\ell',\ell+1} \sqrt{\frac{{(\ell-m+1)(\ell+m+1)}}{(2\ell+1)(2\ell+3)}} \right].
\ee
\end{widetext}
All odd multipole coefficients in the spectrum are imaginary $\A^*_{2K+1,M}=-\A_{2K+1,M}$, and the even are real
$\A^*_{2K,M}=\A_{2K,M}$ for any $K$. The geometric coefficients are symmetric,
$\xi^{L M}_{\ell m;\ell' m'} = \xi^{L M}_{\ell' m';\ell m}$ (one can check this is the case for the dipole above).
The angular correlations of course turn out to be symmetric $\langle a_{\ell m} a^*_{\ell' m' } \rangle=\langle a_{\ell' m'} a^*_{\ell m
}\rangle$, though the primordial spectrum may not be, $\langle \R({\bf x}) \R({\bf x}) \rangle \neq \langle \R({\bf x'})\R( {\bf x}) \rangle$. As shown above, this can be understood by the noncommutative quantum nature of the fields whose fluctuations are responsible for the perturbations.

\subsection{Anisotropic power spectrum}

As found in sections \ref{geometry} and \ref{nc_fields}, the two-point function acquires typically exponential modulations from noncommutative geometry. For clarity, let us focus on an exponential term $e^{i{\bf k}\cdot {\bf r_0}}$ in the following.
This be decomposed using the Rayleigh formula and by expressing the Legendre polynomial $P_\ell$ in terms of sum of products of spherical harmonics:
\ba \label{rayleigh}
e^{i{\bf k}\cdot {\bf r_0}}& =& \sum_{\ell=0}^\infty i^\ell(2\ell+1)j_\ell(kr_0)P_\ell(\hat{k}\cdot\hat{r}_0)
= \\ \nonumber &4&\pi \sum_{\ell=0}^\infty i^\ell j_\ell(kr_0) \sum_{m=-\ell}^{\ell}Y_{\ell m}^*(\Omega_k) Y_{\ell m}^*(\Omega_{r_0})\,.
\ea
Now comparing the spectra with the general form (\ref{primordial}), and using the orthogonality of spherical harmonics together with (\ref{rayleigh}), we obtain the amplitudes of each modulations:
\be \label{modulations}
\A_{LM} = 4\pi i^L j_L(kr_0)Y_{LM}^*(\Omega_{r_0})\A\,.
\ee
Clearly the scale dependence of these coefficients cannot be described by simple power-laws. Instead, the modulations will be oscillating along the $k$-modes.

Let us first comment the power spectrum. we note that already the isotropic spectrum is modified with respect to the usual result
$\A_{00}(\mu=0) \equiv \A$, because of the nontrivial $k$-dependence encoded in the function $j_0(kr_0)=\sin(kr_0)/kr_0$. In principle the oscillatory behavior of the modulation could result in ''wiggles'' there seem to appear in the observed spectrum. Such wiggles have also been predicted from transplanckian physics \citep{Martin:2003kp} or from cyclic inflation \citep{Biswas:2010si}, and the data has been shown to be compatible with such features.

Furthermore, there is an infinite series of higher-multipole modulations which will introduce statistically anisotropic correlations. The reflects the nonlocality of the underlying model. In principle all types of modulations are present, meaning that every $\ell$-mode is coupled to any other. Each wavemode of perturbations that contributes to the power spectrum is also relevant to the anisotropic couplings. In particular, as one expects from ultraviolet noncommutativity, small wavelengths contribute most to the modulations at all scales. Contribution from extremely small wavelengths would cancel out due to rapid oscillations.

It is crucial to note however, that in practise the scales at which the modulations are strongest, do not contribute to the angular modes observed in the CMB unless the length scale of noncommutativity is much above the Planck scale.
The spherical Bessel functions $j_\ell(z)$ have their highest peak at about $z \sim L$, so each $L$-modulation will be strongest at wavemodes corresponding to $k \sim L/r_0$. On the other hand, the contribution to the CMB anisotropy $\Theta_\ell(k)$ from inhomogeneous sources $\Theta(k,\tau)$ is also dependent on the spherical Bessel function
\be \label{source}
\Theta_\ell(k) = \int j_\ell(k\tau)\Theta(k,\tau)d\tau\,.
\ee
The scales contributing most to the multipole $\ell$ are $k \sim \ell/\tau_*$, where the comoving distance to the last scattering surface is
about $\tau_* \approx 14000$ Mpc. Thus the modulation of the order $L$ will affect maximally the correlators corresponding to the multipole $\ell$
when
\be
r_0 \sim 1.4\cdot 10^{61} \lp\frac{L}{\ell}\rp M_P^{-1} \sim H\theta \sim  \frac{H}{\mu^2} \,,
\ee
where $M_P^{-1}$ is the Planck length and $\mu^{-1} \sim \theta^{-\frac{1}{2}}$ is the noncommutative length scale. So, the dipole modulation at the multipoles $\ell \sim 1000$ would be of order one if $\mu \sim 10^{-30}M_P \sim 10^{-10} \Lambda_{QCD}$ at inflation. Now, to translate this into the physical energy scale of noncommutativity observable today in laboratory we should recall the relation (\ref{cm_ph}). If the reheating temperature of the universe was close to the GUT scale $\sim 10^{16}$ GeV, the scale factor at the end of inflation was about $a_{RH} \sim 10^{-29}$ of its value normalized to unity at the present. This gives us $\mu^{ph} \sim (\theta^{ph})^{-\frac{1}{2}} \sim 10^{-16} M_P \sim 10$ TeV.
 We note also that at those multipoles the cosmic variance is negligible and Planck can be expected to measure deviations from statistical anisotropy at percent level or so. Furthermore, the total effect of the modulations does of course not come from the peak of the Bessel functions in (\ref{modulations}) but is the cumulative contribution integrated from all scales. We may then expect several orders of magnitude improvement to the above estimate of the maximal noncommutative scale $\mu$ that may be observed in the CMB. Conservative lower bounds from modifications to standard model of particle physics give $\mu \gtrsim $ few TeV \citep{PhysRevD.64.075012,Mocioiu:2000ip,PhysRevLett.87.141601}. Thus, the tightest bounds may turn out to be cosmological.

 Full comparison with the data would require considerable technical difficulties, firstly because all the observed multipoles should be included in the analysis, and even higher k-modes than usually corresponding to those would have to be taken into account. Moreover, since there occur couplings between arbitrarily separated $\ell$-modes, one cannot employ the previous techniques that have been developed to deal with sparse correlation matrices (with only the diagonal and some adjacent entries nonvanishing). Finally, the distortion of the power spectra should be tested in conjunction with the effects of the anisotropic correlations.

Therefore, and because both observations and theory suggest these effects should be small,
let us then, instead of the full pattern, consider the power series expansion
\be
e^{i{\bf k}\cdot {\bf r_0}} \approx 1 + i{\bf k}\cdot {\bf r_0} - \frac{1}{2}({\bf k}\cdot {\bf r_0})^2 + \dots
\ee
Note that the expansion of the more general parametrization (\ref{geometry_p2}) is essentially very similar.
It is useful to separate the magnitude $r$ of ${\bf r}_0=r\rr$, defining the unit direction vector $\rr$ decomposed as
\be
\rr_\pm = \mp\lp\frac{r_x\mp i r_y}{\sqrt{2}r}\rp\,, \quad \rr_0 = \frac{r_z}{r}\,, \quad r=|{\bf r}_0|\,.
\ee
Then the nonvanishing contributions to the spectrum may be written as the following. The amplitudes are
\ba \label{dipole1}
\A_{00} & = & \A\,,  \quad
\A_{1(-1)}  =  2i\sqrt{\frac{\pi}{3}}k_0 r\rr_{-}\A\,,  \\
\A_{10}  &=&  2i\sqrt{\frac{\pi}{3}}k_0 r\rr_{0}\A\,, \quad\nonumber
\A_{1(+1)}  =  2i\sqrt{\frac{\pi}{3}}k_0 r\rr_{+}\A\,,
\ea
and the corresponding spectral indices are
\be \label{dipole2}
n_{00} = n_s\,, \quad n_{1m} = 1+n_s\,.
\ee
In a companion paper we test the leading order dipole correction given by (\ref{dipole1}) with the data, which is found to slightly prefer the presence of the dipole \citep{nicolaas}.

\subsection{Comparison with imperfect source models}
\label{comparison}

Let us remark on the difference to the imperfect dark energy model, where similar geometric modulations can appear as well.
There the dipole is due to an anisotropic source,
it's contribution with respect to the quadropole is subdominant. This is because there
then occurs cancellation to the odd correlations, as they are proportional to
\ba
\langle a_{\ell m}a^*_{(\ell+L)m'}\rangle_{DE} \sim  \qquad\qquad \qquad\qquad\qquad\qquad  \nonumber \\ A_0\int
d(\log{k})\left[\Theta_\ell(k)\Theta^A_{\ell+L}(k)-\Theta_{\ell+L}(k)\Theta^A_{\ell}(k)\right]\,,
\ea
where $A$ quantifies the magnitude of the anisotropy, $\Theta_\ell^A(k)$ is the anisotropic transfer function and $L$ is odd.
The function (\ref{source}) gives the source contribution in the isotropic case, in the presence of imperfect sources the $\Theta$'s depend also on the direction of the wavevector.
Due to the partial cancellation effect in such a case, \cite{magnus} found the quadropole contribution dominant even though it was suppressed by the small
parameter corresponding to $A_0^2$. However, since in the present case the dipole is of primordial origin, we have
\be
\langle a_{\ell m}a^*_{(\ell+L)m'}\rangle \sim \A_{LM}\int d(\log{k})\Theta_\ell(k)\Theta_{\ell+L}(k)\,,
\ee
and the magnitude of the odd and even modulations is expected to be similar. A priori, higher multipoles are again suppressed by some
small parameter, and the dominating correction to the monopole is now generically the dipole.

\subsection{Non-Gaussianities}
\label{nong}

The non-commutativity also affects the non-Gaussian statistics of
primordial perturbations. Here we discuss non-Gaussianities in
effective single field models, described by (\ref{deltaN}), where
the field $\phi$ affects very little the classical dynamics during
inflation, $V'/3H^2M^2 \sim 0$, but becomes dynamically relevant at
some later stage. Such models can generate observable
non-Gaussianities in the usual commutative case \citep{Enqvist:2004bk,Lyth:2005fi}, see
e.g. works concerning the curvaton scenario \citep{Bartolo:2003jx,Enqvist:2005pg} or modulated
reheating \citep{Enqvist:2004ey,Ichikawa:2008ne}, and our aim is to analyze how the
non-commutativity alters the predictions.

Assuming standard slow roll dynamics with canonical kinetic terms
during inflation, the Fourier transform of (\ref{deltaN}) can
expressed in the form
  \beq
  \label{deltaN_fourier}
  \zeta_{\theta}(t,\k)=\tilde{\zeta}_{\theta}(t,\k)+f(k)\int\frac{\dq}{(2\pi)^3}
  \tilde{\zeta}_{\theta}(t,\q)\tilde{\zeta}_{\theta}(t,\k-\q)+\ldots\
  ,
  \eeq
where we have defined
  \baq
  \label{tilde_zeta}
  \tilde{\zeta}_{\theta}(t,\k)&=&N'\left(1-\eta\,\lnkkp\right)\delta\phi_{\theta}(t_{k},\k)\
  ,\\
  \label{fk}
  f(k)&=&\frac{N''}{2N'^2}\left(1+\nfnl\,\lnkkp\right)\ ,\\
  \label{nf}
  \nfnl&=&\frac{N'}{N''}\left(
  -3\eta+
  \frac{V'''}{3H^2}\right) \ .
  \eaq
Here we have neglected slow roll corrections to constant terms and
derived the $k$-dependent terms to leading order precision in slow
roll. The same precision is used in the results derived below.
$N=N(t,t_i)$ and all other quantities in (\ref{tilde_zeta}),
(\ref{fk}) and (\ref{nf}) without explicit time indices are
evaluated at the time $t_i$ appearing in (\ref{deltaN}). The slow
roll parameter $\eta$ is defined by $\eta=M_{\rm P}^2V''/3H^2$ and
$V$ denotes the potential of $\phi$.

In the commutative case, the field $\tilde{\zeta}$ is Gaussian to
leading order in slow roll. The magnitude of primordial
non-Gaussianities is then controlled by the function $f(k)$ which is
related to the non-linearity parameter $f_{\rm NL}$ for equilateral
configurations. The scale-dependence of $f_{\rm NL}$ is measured by
$n_{f_{\rm NL}}$.

In non-commutative theories, $\tilde{\zeta}_{\theta}$ becomes
non-Gaussian due to the inherent non-Gaussianities of the
fluctuations $\delta\phi_{\theta}$. This affects both the magnitude
of $f_{\rm NL}$ and its scale-dependence. Quite generally, the
non-Gaussianities also deviate from the (quasi-)local form since the
non-Gaussianity in (\ref{tilde_zeta}) is not of the simple
(Gaussian)$^2$ type. Below we analyze non-Gaussianities arising in non-commutative theories discussed in
section \ref{geometry}. (The non-Gaussianity in the approach of section \ref{nc_fields} can be considered starting from (\ref{tpc_h}), but for clarity we do not
consider that separately here).

Using the relations between n-point functions of
$\delta\phi_{\theta}$ and $\delta\phi_{0}$ given in section
\ref{geometry}, we can express the three-point function of
$\zeta_{\theta}$ in the form
  \baq
  \label{threepoint_full}
  \langle\zeta_{\theta}(t,\k_1)\zeta_{\theta}(t,\k_2)\zeta_{\theta}(t,\k_3)\rangle
  =(2\pi)^3\delta(\sum \k_i)& &  \nonumber \\  \times e^{-i\k_1\wedge\k_2} \left(\rule{0pt}{3ex}\right.2P_{0}(k_1)
  P_{0}(k_2)f(k_3)e^{2 H\vec{\theta}\cdot\bar{k}_3} & + & 2\,{\rm
  p.}\left)\rule{0pt}{3ex}\right.\ ,
  \eaq
where $P_{0}=N'^2H^2/2k^3$ denotes the power spectrum in the
commuting case. In this section we denote the wavevectors by an overbar, $\k={\bf k}$.

To obtain real-valued results in coordinate space, we apply the prescription introduced in (\ref{pres}) to (\ref{threepoint_full}) and identify the observable three-point correlator with
\baq
\langle\zeta_{\theta}\zeta_{\theta}\zeta_{\theta}\rangle & \equiv &
\alpha \langle\zeta_{\theta}\zeta_{\theta}\zeta_{\theta}\rangle_{\rm M} +
i(1-\alpha)\langle\zeta_{\theta}\zeta_{\theta}\zeta_{\theta}\rangle_{\rm A} \nonumber \\
& \equiv & (2\pi)^3\delta(\sum
  \k_i)B_{\theta}(\k_1,\k_2,\k_3)\,,
\eaq
where the self-adjoint part is
\baq
\langle\zeta_{\theta}(\k_1)\zeta_{\theta}(\k_2)\zeta_{\theta}(\k_3)\rangle_{\rm M} & = &  \\ \nonumber
\frac{1}{2}\Big( \langle\zeta_{\theta}(\k_1)\zeta_{\theta}(\k_2)\zeta_{\theta}(\k_3)\rangle
& + & \langle\zeta_{\theta}(-\k_1)\zeta_{\theta}(-\k_2)\zeta_{\theta}(-\k_3)\rangle^*\Big)\,,
\eaq
and the antiself-adjoint part is
\baq
\langle\zeta_{\theta}(\k_1)\zeta_{\theta}(\k_2)\zeta_{\theta}(\k_3)\rangle_{\rm A} & = &  \\ \nonumber
\frac{1}{2}\Big( \langle\zeta_{\theta}(\k_1)\zeta_{\theta}(\k_2)\zeta_{\theta}(\k_3)\rangle
& - & \langle\zeta_{\theta}(-\k_1)\zeta_{\theta}(-\k_2)\zeta_{\theta}(-\k_3)\rangle^*\Big)\,,
\eaq
and all perturbations are evaluated at the same time $t$. It becomes then straightforward to derive the
result
\baq \label{bispectrum}
B_{\theta}(\k_1,\k_2,\k_3)  & = &   \\
2e_\alpha^{-i\k_1\wedge\k_2}&\Big(&\cosh(2H\vec{\theta}\cdot\k_3)P_0(\k_1)P_0(\k_2)f_3(\k_3)    +   2\,{\rm  p.}\Big)\nonumber \\
+2i e_{1-\alpha}^{i\k_1\wedge\k_2}&\Big(&\sinh(2H\vec{\theta}\cdot\k_3)P_0(\k_1)P_0(\k_2)f_3(\k_3)  +   2\,{\rm  p.}\Big)\,. \nonumber
\eaq
We used a shorthand notation $+2\,{\rm  p.}$ to denote the permutations of the three indices.
Using the above result and the spectrum of the two-point function computed in
section \ref{geometry}, we find the non-linearity parameter $\fnlth$
given by the expression
  \baq
  \label{fnl}
  f_{{\rm NL},\theta}&\equiv&\frac{5}{6}\frac{B_{\theta}(\k_1,\k_2,\k_3)}
  {P_{\theta}(k_1)P_{\theta}(k_2)+2\,{\rm p.}}\\\nonumber
  &=&\frac{5}{3}\exp_\alpha(-i\k_1\wedge\k_2) \\ \nonumber &\times& \frac{P_{0}(k_1)P_{0}(k_2)f(k_3){\rm cosh}(2H\vec{\theta}\cdot\k_3)+2\,{\rm
  p.}}{P_{0}(k_1)P_{0}(k_2)e_\alpha(H\vec{\theta}\cdot\k_1)e_\alpha(H\vec{\theta}\cdot\k_2)+2\,{\rm p.}}\\\nonumber
  &+&
   i\,\frac{5}{3}\exp_{1-\alpha}(i\k_1\wedge\k_2) \\ &\times& \frac{P_{0}(k_1)P_{0}(k_2)f(k_3){\rm sinh}(2H\vec{\theta}\cdot\k_3)+2\,{\rm
  p.}}{P_{0}(k_1)P_{0}(k_2)e_\alpha(H\vec{\theta}\cdot\k_1)e_\alpha(H\vec{\theta}\cdot\k_2)+2\,{\rm p.}}\,. \nonumber
  \eaq
We have separated the real and imaginary terms in (\ref{bispectrum}) and (\ref{fnl}). The imaginary contributions violate
parity and they vanish if all the components of $\vec{\theta}$
are set to zero. Otherwise they are present for arbitrary $\alpha$: in particular we observe that restriction to the self-adjoint piece of the correlation ($\alpha=1$) does not eliminate the odd correlations. One also expects that higher order correlations would exhibit parity violations even in the case of purely spatial noncommutativity.
The spatial components of the non-commutativity
matrix $\theta_{ij}$ enter the through results the phase ${\rm
exp}(i\k_1\wedge\k_2)={\rm exp}(k_1^ik_2^j\,\theta_{ij})$. They do
not appear in the results for the spectrum and therefore affect only
the non-Gaussian statistics of primordial perturbations.

For simplicity, we analyze in the following only the modifications due to
$\theta_{ij}$ setting all components of $\vec{\theta}$ equal to
zero. This gives
  \baq
  \label{fnl_theta_ij}
  f_{{\rm NL},\theta}
  &=&\frac{5}{3}e_\alpha^{-i\k_1\wedge\k_2}\frac{P_{0}(k_1)P_{0}(k_2)f(k_3)+2\,{\rm
  p.}}{P_{0}(k_1)P_{0}(k_2)+2\,{\rm p.}}\ ,
  \eaq
where the only contribution from the non-commutativity is the prefactor
involving the wedge product. This affects the scale dependence of
$\fnlth$ and can hence be constrained observationally. For example,
computing the scale-dependence for shape preserving variations of
the momentum space triangle, $\k_i\rightarrow \lambda \k_i$,
defined as
\be
n_{\fnlth} =  \frac{\partial {\rm ln}\,|f_{{\rm NL},\theta}(\lambda\k_1,\lambda\k_2,\lambda\k_3)|}{\partial {\rm ln}\,
  \lambda}\Big|_{\lambda=1}\,,
\ee
we find, in the two specific cases,
\be \label{nfnl_result}
n_{\fnlth} = \left\{ \begin{array}{ccc}   2k_1^ik_2^j\theta_{ij}\cot(k_1^ik_2^j\theta_{ij}) +\nfnl & \text{if} & \alpha=0\,,  \\
 -2k_1^ik_2^j\theta_{ij}\tan(k_1^ik_2^j\theta_{ij})+\nfnl
&  \text{if} & \alpha=1\,. \end{array}\right.
\ee
where $\nfnl$ given by (\ref{nf}) is the result in the commuting
case. The part dependent on $\theta_{ij}$ arises purely from
non-commutative features. The observational prospects of scale dependent
$f_{\rm NL}$ were considered in Ref.\citep{Sefusatti:2009xu}, which suggests that
Planck data could be sensitive to a scale dependence of the order of
slow roll parameters. The scale dependence therefore could place
interesting bounds on $\theta_{ij}$. Moreover, it is worth noting
that the result (\ref{nfnl_result}) depends on the wavevectors
$\k_1$ and $\k_2$ and hence on the shape of the momentum space
triangle. This is in contrast with the commutative case, where the
shape dependence is given by the same result $\nfnl$ for all shape
preserving variations, $\k_i\rightarrow \lambda \k_i$, regardless of
triangle shape. This allows, in principle, to distinguish between
the contributions arising from the non-commutative properties of the
theory and from the standard classical inflationary physics.

\section{Conclusions}
\label{conclusions}

We considered the effects of noncommutative geometry to the statistics of the CMB anisotropy field. The results are encoded in two formulas:
\begin{itemize}
\item The statistically anisotropic modulation of the two-point function: Eq.(\ref{geometry_p2}).
\item The function $f_{NL}$ characterizing the non-Gaussian property of the three-point function: Eq.(\ref{fnl}).
\end{itemize}
Both of these describe effects that in general violate parity. The presence of spacetime noncommutativity was found to induce the leading contributions to the anisotropic couplings, which occur in principle between all pairs of multipoles. The non-Gaussianity is scale-dependent in a way which depends upon the shape of the momentum triangles considered. These features can provide stringent bounds on the scale of noncommutativity.

The first tests of these predictions are underway \citep{nicolaas}. A quite promising result is that already the leading contribution, a dipole modulation, is found to have an anomalous signature which exhibits a hemispherical asymmetry and is modestly preferred by the data. We hope to make progress also on the theoretical problem of the physical part of the correlations in a future publication. In particular, the correct value of $\alpha$, introduced in (\ref{pres}), should be deduced from first principles, whereas it was left here as an additional parameter to be determined empirically.

By looking closely at the {\it odd} features in the sky, one may see evidence that an accurate description of the universe must be {\it deformed} and {\it twisted}, since it is fundamentally {\it pointless}.

\section{Acknowledgments}
We are grateful to Sami Nurmi for his invaluable contribution, discussions and comments.
 TK is supported by the Academy of Finland
and the Yggdrasil grant of the Research Council of Norway.  DFM
thanks the Research Council of Norway FRINAT grant 197251/V30 and
the Abel extraordinary chair UCM-EEA-ABEL-03-2010. DFM is also
partially supported by the projects CERN/FP/109381/2009 and
PTDC/FIS/102742/2008.

\section*{}

\bibliography{drefs}

\end{document}